\documentclass[aps, prb, amsfonts, amssymb, tightenlines, reprint, superscriptaddress, a4paper]{revtex4-2}

\usepackage{dcolumn}% Align table columns on decimal point
\usepackage{bm}% bold math
\usepackage{amsmath}
\usepackage[export]{adjustbox}

\usepackage{graphicx}   % need for figures
\usepackage{color}      % use if color is used in text
\usepackage{subfigure}  % use for side-by-side figures
                                          
\usepackage{hyperref}   % use for hypertext links, including those to external documents and URLs

\usepackage{tikz}
\usetikzlibrary{positioning,arrows}
\usetikzlibrary{decorations.pathmorphing}
\usetikzlibrary{decorations.markings}

\newcommand{\bvec}[1]{\mathbf{#1}}

\begin{document}

\title{Interplay of Electron-Magnon Scattering and Spin-Orbit Induced Electronic Spin-Flip Scattering in a two-band Stoner model}

\author{Felix Dusabirane\thanks}
\email{f.dusabirane@ur.ac.rw}
\affiliation{Physics Department and Research Center OPTIMAS, RPTU University of Kaiserslautern-Landau, 67663 Kaiserslautern, Germany}
\affiliation{Department of Physics, College of Science and Technology, University of Rwanda, P.O. Box 3900, Kigali, Rwanda.}
\author{Kai Leckron}
\author{Baerbel Rethfeld}
\author{Hans Christian Schneider\thanks }
\email{hc.schneider@rptu.de}
\affiliation{Physics Department and Research Center OPTIMAS, RPTU University of Kaiserslautern-Landau, 67663 Kaiserslautern, Germany}

\begin{abstract}
This paper presents a theoretical investigation of electron-magnon scattering processes in the ultrafast demagnetization in itinerant ferromagnets. In the framework of a ferromagnetic model system, we compute the spin-dependent dynamics of electrons in itinerant Bloch states by including electron-magnon and electron-electron scattering processes on an equal footing. While the former process flips the electronic spin accompanied by the creation or destruction of a magnon, the latter exchanges electronic angular momentum with the lattice due to the influence of spin-orbit coupling. We show that, for a realistic choice of the electron-magnon interaction and deposited pulse energy, the \emph{interplay} of these two different scattering mechanisms leads to the creation of magnons and a transfer of angular momentum to the lattice that constitutes an essentially non-equilibrium microscopic scenario for the ultrafast demagnetization process in itinerant ferromagnets.
\end{abstract}

\maketitle

\section{Introduction}

The discovery of optically induced ultrafast demagnetization in metallic ferromagnets~\cite{beaurepaire1996ultrafast} has led to the question how the spin angular momentum of excited electrons is dissipated on a 100\,fs timescale. Most theoretical proposals that deal with local contributions involve spin-orbit coupling in one way or another, but there may also be nonlocal contributions from spin-dependent transport~\cite{battiato2010superdiffusive,nenno2018particle}.  It was suggested early on that a mechanism of the Elliott-Yafet (EY) type could be responsible for the angular momentum dissipation, either to phonons or to the rigid lattice~\cite{koopmans2005unifying,koopmans2010explaining,walowski2008energy}. This mechanism describes how even spin-independent scattering processes, such as the electronic Coulomb scattering and the spin-independent part of the electron-phonon interaction, can change the electronic spin in the presence of spin-orbit coupling. The main purpose of this paper is to start from the picture of EY-like spin-changing transitions, where effectively the lattice acts as a spin sink, and include electronic scattering processes with magnons. We intend to show that the interplay of these two electronic scattering processes, which both change the electronic spin in different ways, leads to ultrafast demagnetization dynamics involving an effective angular momentum exchange with the lattice that is more efficient than either of these processes acting alone.

\section{Demagnetization Mechanisms}
\label{sec:demag}

We would like to present here an overview of different possible contributions to the ultrafast demagnetization dynamics in ferromagnets. It seems necessary to provide some background and explain our approach as no comprehensive picture of this process exists, despite the field being 30 years old by now. In particular, no consensus has been reached as to how the sizable angular momentum from the quenched magnetization is transferred on ultrashort timescales and beyond. 

When classifying possible mechanisms for ultrafast demagnetization or comparing calculations for different candidate mechanisms, key questions include (1) on what time scales the magnetization is affected and (2) which deposited pulse energy is needed to achieve experimentally observed magnitudes of the demagnetization in metallic ferromagnets. As far as intrinsic time scales are concerned, the electronic dynamics are influenced directly by the electromagnetic field during the first few femtoseconds. In this regime, which may be described at the level of electronic quantum-state dynamics~\cite{Toews-prl,Zhang-2000} or by time-dependent density-functional theory (DFT)~~\cite{elliott_microscopic_2020,dewhurst_angular_2021}, the electromagnetic field coherently couples to the electronic orbital degrees of freedom, and via the spin-orbit interaction also affects the spin. During the first few hundred femtoseconds one expects that dephasing and scattering processes with a large available phase space establish a pre-equilibrated state that is governed by electronic distributions, which still may be far from equilibrium. After around a picosecond it is likely that a quasi-equilibrium regime characterized by local temperatures and chemical potentials is reached. Remagnetization, i.e., relaxation back to a complete equilibrium proceeds mainly on the time scale of several picoseconds and beyond.

The direct and coherent influence of optical fields with very short pulse durations and very high peak fields can induce remarkable effects, such as sub-femtosecond control of currents~\cite{higuchi2017light} and magnetism~\cite{siegrist2019light}. On these time scales, one finds a pronounced influence of the details of the driving field on the electronic dynamics. However, the demagnetization dynamics, as typically driven by pulses on the order of several 10 femtoseconds and recorded by the magneto-optical Kerr effect~\cite{beaurepaire1996ultrafast}, happen on a time scale longer than that of the just mentioned purely coherent effects. 
Therefore the magnetization may be affected by the coherent influence of the optical field on the electronic wavefunctions and/or by scattering and dephasing dynamics during which the phase and the influence of the pulse details are washed out. This question was studied recently by Stiehl and coworkers~\cite{stiehl_role_2022} who refer to the coherent influence on the magnetization as a ``primary'' effect and other incoherent contributions as ``secondary'' effects. In Ref.~\cite{stiehl_role_2022} the influence of the pump photon energy on the excited carrier densities in nickel was compared with the measured demagnetization dynamics, which were normalized to the deposited energy. It was found that the demagnetization traces are determined by the energy deposited by the drive pulse and not by the details of the excited electron distributions. This is a strong indication that the demagnetization process is indeed mainly due to secondary effects, and the pulse excitation may be modeled by incoherent electron distributions in the ferromagnet after the phase and memory of the coherent drive pulse are lost. This view is also supported by a recent time-dependent DFT study~\cite{mrudul_ab_2024}, which reported the deposited energies necessary for different demagnetization dynamics. It turns out that in order to obtain an impact on the demagnetization that is comparable to typical experimental results, their calculation of the coherent demagnatization contribution, i.e., the ``primary'' effect,  needs an unrealistically high deposited energy. More specifically it was found that in a ferromagnetic alloy, a sizable quenching in the magnetization due to the coherent excitation process alone is only obtained if the pulse deposits an energy of 3\,eV per unit cell~\cite{mrudul_ab_2024}. This is much larger than the estimate of about 150\,meV per unit cell, which was obtained in Ref.~\cite{essert_electron-phonon_2011} based on typical experimental fluences~\cite{stiehl_role_2022}. We therefore do not consider here the influence of the coherent excitation process~\cite{elliott_microscopic_2020,dewhurst_angular_2021} and focus on incoherent scattering processes.

Next in the progression of time scales are processes that affect the magnetization via a change of the ensemble electron spin due to scattering processes. The potentially relevant \emph{incoherent} electronic interaction mechanisms are electron-phonon, electron-electron and electron-magnon interactions. For the quenching of the magnetization in ferromagnets, electron-electron and electron-phonon interactions have mostly been studied in connection with spin-orbit coupling which makes it possible even for explicitly spin-independent interactions to transfer electronic-spin angular momentum to the lattice. Such a description in terms of electronic transition rates due to phonon absorption and emission at the level of Fermi's Golden rule, sometimes referred to as EY mechanism, has been successfully applied to the spin relaxation in materials with spin degenerate bands, i.e., semiconductors and simple metals~\cite{yafet_g_1963,fabian_spin_1998,grimaldi_theory_1997,baral_re-examination_2016,vollmar_generalized_2017}. For electron-phonon scattering in these systems, it is important to consider both the explicitly spin-dependent and spin-independent contributions to the matrix elements and the transition probabilities. As far as the electron-phonon interaction has been analyzed for ferromagnets~\cite{steiauf_elliott-yafet_2009,essert_electron-phonon_2011,carva_ab_2011}, the spin-relaxation/demagnetization times have been in agreement with experiments, but the achievable demagnetization for realistic fluences was too small. Further, the main contribution to this process comes from longitudinal phonons via the explicitly spin-independent part of the interaction~\cite{essert_electron-phonon_2011}. As the longitudinal phonons are characterized by zero ``phonon spin'' angular momentum, there has been no explicit contribution to the angular-momentum balance from angular momentum carried by phonons in these calculations. Further, the EY mechanism is also effective for the electron-electron interaction~\cite{krauss2009ultrafast,mueller_feedback_2013}, which typically leads to an even faster dynamics in highly excited electron systems than the electron-phonon interaction. As can be checked by including spin coherences due to internal effective spin-orbit fields and their dephasing, i.e., Dyakonov-Perel like dynamics~\cite{leckron_ultrafast_2017,vollmar_ultrafast23} or correlation effects~\cite{Faehnle-marginal-17}, the transfer of angular momentum due to spin-independent scattering processes in ferromagnets can be well described at the level of incoherent transition rates. These lead to demagnetization time scales~\cite{steiauf_elliott-yafet_2009,essert_electron-phonon_2011,carva_ab_2011,krauss2009ultrafast,mueller_feedback_2013} that are also in quantitative agreement with experiments. The incoherent spin-dependent scattering dynamics at the level of momentum or energy-dependent transitions rates can be well understood as being driven by the difference of spin-dependent chemical potentials~\cite{mueller_feedback_2013}. However, the quenching in this demagnetization mechanism depends sensitively on the spin-orbit coupling parameters and assumptions on the excitation induced band-structure change~\cite{mueller_feedback_2013}. 
 
The recent discovery of polarized phonons on time scales relevant for the demagnetization process~\cite{tauchert_polarized_2022} has revealed a new piece of the ultrafast-demagnetization puzzle. It now seems possible that the angular momentum carried by transverse phonons may contribute to the angular momentum transfer involved in the demagnetization process. This question is connected to the rapidly growing interest in phonon angular momentum and phonon chirality in the solid state~\cite{Juraschek-review-2025}. Most of the theoretical work on the coupling of phonon and spin  angular momentum for transverse phonons is done in the framework of localized-spin models~\cite{nakane_angular_2018,ruckriegel_angular_2020}. In these effective theories there is no need for a microscopic coupling of the polarized phonon modes to electrons. While it has been shown how polarized phonons can arise dynamically due to their coupling to magnons~\cite{weissenhofer2024arxiv-trulychiral} and how the rotational degrees of freedom can be driven on ultrashort timescales by coherent effects~\cite{mrudul2025arxiv}, it is not clear at present how to quantify the transfer of phonon orbital momentum to electrons at the microscopic level of interacting electron-phonon dynamics. This task is much easier for the coupling between electrons and magnons which can be modeled as an exchange interaction. As exchange scattering processes can act on ultrashort timescales far away from equilibrium~\cite{baral-exchange}, it seems promising to investigate the contribution of electron-magnon scattering to ultrafast demagnetization and the angular momentum balance. A step towards addressing the requirements of time scale, quenching and fluence  for this mechanism is provided by the present paper by numerically computing the electronic dynamics in a model system due to the interplay of electron-magnon scattering and EY-like scattering processes. From the calculation of the interacting electron and magnon dynamics we show how the loss of magnetization at short times can arise as a non-equilibrium process due to the interplay of the different electronic scattering mechanisms. Furthermore, we argue that an additional relaxation mechanism for the magnons may be needed in order to describe the remagnetization process. 

The paper is organized as follows. In Sec.~\ref{methodology} we describe our general approach and the detailed model of electrons, magnons and their interactions in a band structure with two Stoner spin-split bands. We set up the dynamical equations for the distribution functions of electrons and magnons including the relevant scattering mechanisms; for electron-electron scattering we include the effect of spin-orbit coupling by an effectively spin-dependent interaction matrix element. In Sec.~\ref{Results_and_discussion} we calculate results for the dynamics of the distribution functions after an instantaneous excitation process. We show that magnons at high energies and~$q$ vectors are efficiently created by electron-magnon interactions and highlight the impact of EY-like spin-flip scattering of the electrons. We also show the importance of different relaxation processes for the remagnetization dynamics and present our conclusions in Sec.~\ref{conclusion}.

\section{Model and Dynamical Equations}
\label{methodology}

\subsection{General Approach}

The importance of magnons for ultrafast demagnetization dynamics in ferromagnets has been stressed repeatedly in the literature~\cite{carpene_dynamics_2008,manchon_theory_2012,haag_role_2014,turgut2016stoner,eich_band_2017}. On the basis of separate electron-phonon and electron-magnon transition rates, the interplay of electron-magnon and electron-phonon scattering was suggested as a candidate process for ultrafast demagnetization by F\"{a}hnle and coworkers~\cite{illg2013ultrafast}, and the present paper continues this line of investigation. 
Recently, the magnon contribution to demagnetization has received renewed attention. For instance, the interaction of electrons with magnons and spin-orbit assisted EY-spin-flips has been included in the derivation of macroscopic transport equations for ferromagnet-metal heterostructures~\cite{beens_s-d_2020,beens_modeling_2022}, and a $(N+2)$-temperature model including mode-resolved magnon temperatures has been introduced~\cite{Weissenhofer_Oppeneer_2024}. 

In contrast to Refs.~\cite{beens_s-d_2020,Weissenhofer_Oppeneer_2024}, we explore the coupled electron and magnon dynamics during the demagnetization process in a model ferromagnet with the focus on electronic and magnonic non-equilibrium effects. Our focus on the coupled dynamics entails that we compute the time-dependent electronic distribution functions for the itinerant electrons of the ferromagnet instead of assuming Fermi-Dirac distributions at all times. Our approach is suitable to describe the interplay of the two scattering mechanisms, which is particularly important away from equilibrium during the early stages of the demagnetization process. More precisely, we compute the dynamical changes of the electronic distributions by including both of these interaction mechanisms on an equal footing at the level of Boltzmann scattering integrals, extending our earlier work~\cite{krauss2009ultrafast,mueller_feedback_2013}. 

\subsection{Basic Hamiltonian}

In this section, we set up a model and describe our theoretical approach to calculate the influence of electron-magnon, electron-electron and electron-phonon interactions on the ultrafast spin dependent electron dynamics. 

Electrons, magnons and phonons are governed by the following Hamiltonian
\begin{equation} 
	H=H_{\text{e}} + H_{\text{m}} + H_{\text{p}}+ V_{\text{e-e}} + V_{\text{e-m}} + V_{\text{e-p}} + V_{\text{m-p}}
    \label{eq:basic-H}
\end{equation}
which includes the electron-electron Coulomb interaction, the electron-magnon interaction, the electron-phonon interaction and magnon-phonon interactions, respectively. We will treat the first two of these interactions dynamically and the latter two using relaxation-time approximations. The first term describes the single-electron contribution in our model
\begin{equation}
	H_{\text{e}}=\sum_{\bvec{k},\sigma} \epsilon_{\bvec{k},\sigma}c^\dagger_{\bvec{k},\sigma}c_{\bvec{k},\sigma}
\end{equation} 
where we have denoted by $c_{\bvec{k},\sigma}^\dagger$  ($c_{\bvec{k},\sigma}$) the creation (annihilation) operators for electrons with crystal momentum $\bvec{k}$ and spin $\sigma$. The single-electron energies~$\epsilon_{\bvec{k},\sigma}$ may include self-energy effects such as a \emph{dynamical} spin splitting. An explicit form for the energy will be given in Sec.~\ref{subsec:energy-magnetization} further below. An extension to the multi-band case is straightforward, but would enormously complicate the numerical problem. The free-magnon contribution is
\begin{equation}
	H_{\text{m}} =\sum_{\bvec{q}}\hbar\omega_{\bvec{q}}a^\dagger_{\bvec{q}}a_{\bvec{q}}
\end{equation}
where $a_\bvec{q}^\dagger$ ($a_\bvec{q}$) are the creation (annihilation) operators for magnons on the ferromagnetic ground state with wave vector $\bvec{q}$ and energy $\hbar \omega_q$, which we take to be
\begin{equation}
    \hbar\omega(q) = D q^2
    \label{magnon_energy}
\end{equation} 
with the magnon stiffness constant~$D$ and $q=|\bvec{q}|$.

The electron-electron interaction is given by the standard expression  
\begin{equation}
	\begin{split}
 	V_{\text{e-e}} &= \frac{1}{2}\sum_{\bvec{k}_1\bvec{k}_2}
 	{\sum_{\bvec{k}'_1}}' v(\bvec{k}'_1-\bvec{k}_1) \\
 	& \qquad \sum_{\sigma_1,\sigma_2} c^\dagger_{\bvec{k}'_1,\sigma_1}c^\dagger_{\bvec{k}_2 + \bvec{k}_1-\bvec{k}'_1,\sigma_2} 
 	c_{\bvec{k}_2,\sigma_2}c_{\bvec{k}_1,\sigma_1} 
 	\label{eq:V-e-e}
 	\end{split}
 \end{equation}
where the prime on the second sum indicates that the $\bvec{k}'_1 =\bvec{k}_1$ contribution is taken out and
\begin{equation}
    v(\bvec{k}' -\bvec{k}) = \frac{e^2}{\mathcal{V}\varepsilon_0}\frac{1}{|\bvec{k}'-\bvec{k}|^2}
\label{eq:v_q}
\end{equation} 
is the Coulomb potential with the normalization volume~$\mathcal{V}$ and vacuum dielectric constant~$\varepsilon_0$. For the treatment of Coulomb scattering and screening, we will follow Refs.~\cite{bonitz1996numerical,haug2004quantum,vollmar_ultrafast23}.

\subsection{Magnons and Electron-Magnon Interaction}

In the following we derive and subsequently solve a dynamical description of electron-magnon scattering. We separate the interacting ferromagnetic electron system in electronic single-particle states and their collective excitations, i.e., the magnons. Following Edwards~\cite{edwards_paramagnetic_1982}, we introduce separate dynamical degrees of freedom for electrons and magnons by coupling band electrons to a Heisenberg model of localized spins, without assuming that the single-particle states correspond to quasi-free $s$-electrons and that the magnons arise as excitations of the more localized $d$-electrons, as in Refs.~\cite{hong_theory_1999,Weissenhofer_Oppeneer_2024}. 
The introduction of the localized spins is regarded as a formal way to capture the electronic properties related to collective dynamics~\cite{edwards_paramagnetic_1982} in order to avoid the considerable complications of a theory in which both single-particle and collective properties are determined from the interaction between $d$-electrons~\cite{zhukov_gw_2005,schmidt_ultrafast_2010,paischer2023nonlocal}. Since we are interested in the far-from-equilibrium case, we do not use close-to-equilibrium response functions of the interacting electron system~\cite{hertz_fluctuations_1974} or assume that the electronic system is in a quasi-equilibrium at all times during ultrafast demagnetization~\cite{manchon_theory_2012,beens_s-d_2020,Weissenhofer_Oppeneer_2024}. Once we have introduced the Heisenberg model we can go through a microscopic derivation using the coupling parameters of the model as in earlier treatments, which considered a spin system formed by well-localized electronic wave functions~\cite{davis_electron-magnon_1967,white1968magnon,woolsey1970electron}.

We next  present a short version of the arguments of Refs.~\cite{davis_electron-magnon_1967,white1968magnon,woolsey1970electron}. We write
\begin{equation}
        H_{\text{spin}} =- J \sum_{\bvec{R}} \bvec{S}(\bvec{R})\cdot \bvec{S}(\bvec{R}+\boldsymbol{\delta})
 \label{H_spin}
\end{equation}
for the collective degrees of freedom, formed by a localized spin system with spins $\bvec{S}(\bvec{R})$ at lattice sites $\bvec{R}$ with exchange constant~$J$ between neighboring spins. The interaction between these degrees of freedom and those associated with the itinerant electrons is taken to be
\begin{equation}
	H_{\text{e-spin}}=-I \Omega \int \bvec{S}(\bvec{x})\cdot\bvec{s}(\bvec{x})d^3 x,
 \label{eq:H-e-spin}
\end{equation}
where, following Ref.~\cite{woolsey1970electron}, we introduced a different exchange coupling constant~$I$ between the itinerant and localized spin densities. This is treated as a parameter in the present approach but can, in principle, be determined \emph{ab initio}. We will assume  a value of 0.9\,eV, on the same order of magnitude as used by Beens et al.~\cite{beens_s-d_2020}.
In Eq.~\eqref{eq:H-e-spin} we have introduced the volume of a unit cell $\Omega \equiv \mathcal{V} /\mathcal{N}$ with the number of sites~$\mathcal{N}$. For completeness we note the relation of the localized spins $\bvec{S}(\bvec{R})$ to the localized spin density $\bvec{S}(
\bvec{x})=\sum_{\bvec{R}}\bvec{S}(\bvec{R})\delta (\bvec{x}-\bvec{R})$. In Ref.~\cite{woolsey1970electron} the authors were interested in static renormalization effects, such as the shift of the band edge, which can be taken into account in the future, but for now we focus exclusively on the scattering dynamics between unrenormalized Bloch states.
We employ the usual assumption that the spin density of the itinerant electrons 
\begin{equation}
	\bvec{s}(\bvec{x})=\frac{1}{\mathcal{V}}\sum_{\bvec{k},\bvec{k}'}\sum_{\sigma,\sigma'} \langle\bvec{k}\sigma|\hat{\boldsymbol{s}}|\bvec{k}'\sigma'\rangle e^{i(\bvec{k}-\bvec{k}')\cdot\bvec{x}} c^\dagger_{\bvec{k},\sigma} c_{\bvec{k}',\sigma'}
\end{equation}
interacts with the localized spins at lattice sites $\bvec{R}$.
The Holstein-Primakoff transformation approximates the Fourier transformations of the localized spin operators 
\begin{equation}
	\mathbf{S}(\bvec{R}) = \frac{1}{\sqrt{\mathcal{N}}} \sum_{\bvec{q}} \mathbf{S}(\bvec{q}) \exp(i\bvec{q}\cdot\bvec{R})
\end{equation}
by $S_+(\bvec{q}) =  S_x + i S_y = a_{\bvec{q}}$, $S_-(\bvec{q}) = S_{+}^{\dagger}= a^{\dagger}_{-\bvec{q}}$ and
 $S_z(\bvec{q}) = S - a^\dagger_{\bvec{q}} a_{\bvec{q}}$. The resulting Hamiltonian in second quantization is
\begin{equation}
    V_{\text{e-m}}= - \frac{1}{\sqrt{\mathcal{N}}}\sum_{\bvec{q} \bvec{k}} M_\text{e-m}\Big( c_{\bvec{k}  +\bvec{q}, \downarrow}^\dagger c_{\bvec{k}, \uparrow}  a_{\bvec{q}}  +  c_{\bvec{k} - \bvec{q}, \uparrow}^\dagger c_{\bvec{k}, \downarrow}     a_\bvec{q}^\dagger  \Big)
    \label{eq:V-e-m}
\end{equation}
where  $M_{\text{em}}= \frac{I}{2}$ is the matrix element for electron-magnon scattering.

The Hamiltonian~\eqref{eq:V-e-m} implies that electron-magnon scattering events are accompanied by a spin-flip in the electronic system, as illustrated in Fig.~\ref{fig:electron-magnon_scattering_image}.
\begin{figure}[tbh!]
    \centering
    \tikzset{
    particle/.style={thick,draw=blue, postaction={decorate},  decoration={markings,mark=at position .6 with {\arrow[blue]{triangle 45}}}},
    boson/.style={-latex,decorate, draw=red, decoration={snake, segment length=5pt, amplitude=3.0pt, pre length=.1cm, post length=.20cm}},
     }
    \begin{tikzpicture}[node distance=1.0cm and 1.3cm]
    \coordinate[label=left:$\boldsymbol{q}$] (e1);
    \coordinate[below right=of e1] (aux1);
    \coordinate[below left=of aux1,label=left:{\textbf{k},$\uparrow$}] (e2);
    \coordinate[right=1.80cm of aux1, label=:{\textbf{k}+\textbf{q},$\downarrow$}] (aux2);
    \coordinate[right=1.3cm of aux2,  label=left: {\textbf{k},$\downarrow$}] (aux3);
    \coordinate[right=1.80cm of aux3, label=right:] (aux4);
    \coordinate[above right= of aux4,label={\textbf{k}$-$\textbf{q},$\uparrow$}=:{\textbf{k}$-$\textbf{q},$_\uparrow$}] (e3);
    \coordinate[below right=of aux4,label=right:$\boldsymbol{q}$] (e4);
    \draw[boson] (e1) --node {} (aux1);
    \draw[particle] (aux1) -- node {} (aux2);
    \draw[particle] (e2) -- (aux1);
    \draw[particle] (aux3) -- (aux4);
    \draw[particle] (aux4) -- (e3);
    \draw[boson] (aux4) -- (e4);
    \end{tikzpicture}
     \caption{Electronic spin-flip scattering processes due to magnon absorption (left) and magnon emission (right).} 
    \label{fig:electron-magnon_scattering_image}
\end{figure}
In a magnon-absorption vertex shown on the left of Fig.~\ref{fig:electron-magnon_scattering_image}, an incoming spin-up (``$\uparrow$'') electron with wave vector $\bvec{k}$ plus an incoming magnon with wave vector $\bvec{q}$ are connected to an outgoing spin-down (``$\downarrow$'') electron with wave vector $\bvec{k}+\bvec{q}$. Conversely, in a magnon-emission vertex shown on the right of Fig.~\ref{fig:electron-magnon_scattering_image} an incoming spin-down electron with wave vector $\bvec{k}$ is connected to an outgoing spin-up electron with wave vector $\bvec{k}-\bvec{q}$ and an outgoing magnon with wave vector $\bvec{q}$. Since the magnon contributes $-\hbar$ to the angular momentum balance, the total angular momentum is conserved at each vertex and thus in all transitions governed by this interaction.

Processes in which a magnon is emitted by a spin-up electron or in which a magnon is absorbed by a spin-down electron are not shown. These also need spin-orbit coupling, and are sometimes called anti-Stoner processes. They may contribute to minority-spin state renormalizations in weak ferromagnets~\cite{paischer2024anti-Stoner,usachov2024unveiling}, but their contribution to scattering rates in 3d ferromagnets were found to be small~\cite{haag_role_2014}. We neglect them in the present paper, as we are interested in the dynamics dominated by the phase space of excited electrons. In the framework of a more general theory that includes both dynamics and spectral effects such as band renormalizations, it is possible that their contribution may play a more prominent role.

The fundamental equations of motion in our approach are for the distribution functions of electrons $n_{\bvec{k},\sigma}$ and magnons $N_{\bvec{q}}$. A derivation using the dynamical truncation of the hierarchy of equations of motion, which constitutes a general method for the derivation of coupled dynamical equations for the relevant correlation functions~\cite{rossi_theory_2002,kira_semiconductor_2012}, is presented in Appendix~\ref{appendix}. In this approach, incoherent transitions are described at the scattering level for the distributions of electrons and magnons. Denoting  the ensemble average in the Heisenberg picture by $\langle \ldots \rangle$, these are defined as $n_{\bvec{k},\sigma}=\langle c^{\dagger}_{\bvec{k},\sigma} c_{\bvec{k},\sigma}\rangle$, with $\sigma=\uparrow$, $\downarrow$, and $N_{\bvec{q}}=\langle a^\dagger_{\bvec{q}} a_{\bvec{q}}\rangle$. The equations of motion then read
\begin{widetext}
    \begin{align}
        \frac{\partial}{\partial t}n_{\bvec{k},\uparrow}\Big|_{\text{e-m}}& = \frac{2 \pi}{\mathcal{N} \hbar}  \sum_\bvec{q}  M_\text{e-m}^2 \delta(\epsilon_\bvec{k}^\uparrow -  \epsilon_{\bvec{k}+\bvec{q}}^\downarrow  + \hbar\omega_\bvec{q} ) [  n_{\bvec{k}+\bvec{q},\downarrow} (1-n_{\bvec{k},\uparrow})(1 + N_\bvec{q })  -  N_\bvec{q}  n_{\bvec{k},\uparrow}(1-n_{\bvec{k}+\bvec{q},\downarrow}) ] 
        \label{e-m_n_k_up}
        \\
         \frac{\partial}{\partial t}n_{\bvec{k},\downarrow}\Big|_{\text{e-m}}& =\frac{2 \pi}{\mathcal{N} \hbar}  \sum_\bvec{q}  M_\text{e-m}^2 \delta ( \epsilon_\bvec{k}^\downarrow -\epsilon_{\bvec{k}-\bvec{q}}^\uparrow  -\hbar\omega_\bvec{q})[N_\bvec{q}   n_{\bvec{k}-\bvec{q},\uparrow} (1- n_{\bvec{k},\downarrow}) - n_{\bvec{k},\downarrow} (1+N_\bvec{q}) (1 -n_{\bvec{k}-\bvec{q},\uparrow})  ] 
        \label{e-m_n_k_down}
        \\
        \frac{\partial}{\partial t}N_{\bvec{q}}\Big|_{\text{e-m}}& = \frac{2 \pi}{\mathcal{N} \hbar}  \sum_\bvec{k}  M_\text{e-m}^2 \delta( \epsilon_{\bvec{k}+\bvec{q}}^\downarrow - \epsilon_\bvec{k}^\uparrow   - \hbar\omega_\bvec{q} ) [  n_{\bvec{k}+\bvec{q},\downarrow} (1- n_{\bvec{k},\uparrow})(1 + N_\bvec{q })  -  N_\bvec{q}  n_{\bvec{k},\uparrow}(1-n_{\bvec{k}+\bvec{q},\downarrow}) ] 
        \label{df_em}
    \end{align}
\end{widetext}
This system of dynamical equations contains in-scattering and out-scattering rates formally similar to those obtained from Fermi's Golden Rule~\cite{illg2013ultrafast,beens_magnons_2018,Weissenhofer_Oppeneer_2024}, but the rates are calculated for the time-dependent electron and magnon distributions. Equations~\eqref{e-m_n_k_up}--\eqref{df_em}, in combination with the other processes detailed below, can therefore describe far-from-equilibrium situations. It is possible to consistently go beyond the scattering-level approximation by considering renormalization effects and electron-magnon coherences as well as including higher order terms and magnon-phonon coupling~\cite{Weissenhofer_Oppeneer_2024}, but these extensions will be left to future studies.

\subsection{Electron-Electron Interaction}

For metals and systems with itinerant electrons in general, electron-electron scattering processes are important on ultrashort timescales, such as those of the demagnetization process, and we therefore need to include their effect~\cite{zhukov_gw_2005, bonitz1996numerical}. Since the coupling to the lattice plays an important role as a sink of angular momenmtum in connection with Coulomb and phonon scattering, we assume for the electron-electron scattering in the following that the Bloch states are not pure spin states because of spin-orbit coupling, but still label them by their largest spin component~$\sigma$. The interaction vertex for this case is shown in Fig.~\ref{fig:electron-electron_scattering_image}.

At the level corresponding to our treatment of electron-magnon scattering, one finds that the spin dependent equation of motion due to electron-electron interaction for the distribution functions $n_{\bvec{k}\sigma}$ at wave vector $\bvec{k}$ in band $\sigma$  can be expressed in the form~\cite{krauss2009ultrafast}
\begin{widetext}
	\begin{equation}
		\begin{split}
			\frac{d}{dt}n_{\bvec{k},\sigma} \Big|_{\text{e-e}} = & \frac{2\pi}{\hbar}
			\sum_{\sigma'\sigma_1 \sigma'_1} \sum_{\bvec{k_1}\bvec{k}'}  \delta(\epsilon_\bvec{k}^\sigma+\epsilon_{\bvec{k}_1}^{\sigma_1}-\epsilon_{\bvec{k}'}^{\sigma'}-\epsilon_{\bvec{k}_1+\bvec{k}-\bvec{k}'}^{\sigma_1'}) ~W_{|\bvec{k}-\bvec{k}'|}(\sigma,\sigma_1,\sigma', \sigma_1')^2 \\
            & \times    \big[    n_{\bvec{k}_1+\bvec{k}-\bvec{k}',\sigma_1'} n_{\bvec{k}',\sigma'} (1-n_{\bvec{k}_1,\sigma_1}) (1-n_{\bvec{k},\sigma}) 
            -
            (1-n_{\bvec{k}_1+\bvec{k}-\bvec{k}',\sigma_1'}) (1-n_{\bvec{k}',\sigma'})          n_{\bvec{k}_1,\sigma_1}n_{\bvec{k},\sigma}\big]   .
		\end{split}
		\label{df_ee}
	\end{equation}
\end{widetext}
Here we have approximated the dynamically screened Coulomb interaction by its static limit, that is, we employ the screened Coulomb matrix element
\begin{equation}
	\begin{split}
	W_{|\bvec{k}'-\bvec{k}|}(\sigma,\sigma_1,&\sigma', \sigma'_1) = \frac{v(|\bvec{k}'-\bvec{k}|)}{\varepsilon(\bvec{k}'-\bvec{k})}\\
 &\times\begin{cases} \phantom{1}1,\quad \sigma=\sigma' \text{ and } \sigma_1=\sigma_1' \\ 
 \phantom{1}\alpha,\quad \sigma\neq \sigma' \text{ or   } \sigma_1\neq\sigma'_1  \\ 
 \alpha^2, \quad \sigma\neq\sigma' \text{ and } \sigma_1\neq\sigma'_1 \end{cases}.
 \end{split}
 \label{eq:W-sigma}
\end{equation}
where $v$ is defined in Eq.~\eqref{eq:v_q}. This expression includes an effective EY spin-flip factor $\alpha$~\cite{steiauf_elliott-yafet_2009,krauss2009ultrafast, illg2013ultrafast, long2013spin} to account for the overlap of non-pure spin states. If one wants to avoid dealing with spin-orbit coupling in the basic Hamiltonian~\eqref{eq:basic-H} and the full spin density matrix, as included in Ref.~\cite{leckron_ferromagnetic_2019}, the $\alpha$ factor can give a good effective description of spin-flip processes in ferromagnets due to the interplay of electron-electron (or electron-phonon) interactions and spin-orbit coupling~\cite{leckron_ultrafast_2017,vollmar_ultrafast23}.
We choose $\alpha= 0.1$, which, in principle, can be determined \emph{ab initio}, but is also to be regarded as a parameter here. The dimensionless dielectric function is $\varepsilon(|\bvec{k}-\bvec{k}'|)^{-1}=1+\kappa^2/
|\bvec{k}-\bvec{k}'|^2$ where $\kappa$ is the screening parameter, which we estimate according to the Thomas-Fermi model by $\kappa = \sqrt{e^2/\varepsilon_0\;\partial n_{
\text{e}}/\partial \mu}$, where $n_\text{e}$ is the electron density.
The contribution
$\delta(\epsilon_\bvec{k}^\sigma+\epsilon_{\bvec{k}_1}^{\sigma_1}-\epsilon_{\bvec{k}'}^{\sigma'}-\epsilon_{\bvec{k}_1+\bvec{k}-\bvec{k}'}^{\sigma_1'})$ guarantees the energy conservation for this two-electron scattering process~\cite{bonitz1996numerical}. Note that these scattering processes conserve the total number of electrons, total electron energy and total electron momentum, since we exclude umklapp scattering. The total electron spin is not conserved, as we consider Coulomb scattering with nonpure spin states, so that the lattice acts as a spin sink when the electronic spin state is changed in an incoherent transition from an initial to a final state.

\tikzset{
	electron1/.style={thick, draw=blue , postaction={decorate},decoration={markings,mark=at position .7 with {\arrow[blue]{triangle 45}}}},	
    electron2/.style={thick, draw=blue , postaction={decorate},decoration={markings,mark=at position .7 with {\arrow[blue]{triangle 45}}}},
	interaction/.style={thick,  draw=black, decoration={coil,aspect=0, segment length=16pt}, decorate}}
 
\begin{figure}
	\centering
	\begin{tikzpicture}[scale=1]	
		\newcommand{\base}{440pt}
		\newcommand{\xstart}{0.025*\base};
		\newcommand{\ystarta}{-0.1*\base};
		\newcommand{\dx}{0.1*\base};
		\newcommand{\dy}{0.05*\base};
		
		\newcommand{\ystartb}{\ystarta+1.85*\dy}
		
		% --- oben links ---
		\coordinate[] (ea1)	at (\xstart-\dx,\ystarta-\dy);
		\coordinate[] (xa)	at (\xstart    ,\ystarta);
		\coordinate[] (ea2)at (\xstart+\dx,\ystarta-\dy);
		\coordinate[] (eb1)	at (\xstart-\dx, \ystartb+\dy);
		\coordinate[] (xb)	at (\xstart    , \ystartb);
		\coordinate[] (eb2)	at (\xstart+\dx, \ystartb+\dy);
		\draw[electron1] (ea1) -- node[label=below left:$\left|\mathbf{k}\mathrm{,}\sigma\right\rangle$,yshift=-5pt] {}(xa);
		\draw[electron2] (xa) -- node[label=below right:$\left|\mathbf{k}'\mathrm{,}\sigma'\right\rangle$, xshift=0pt, yshift=-5pt] {}(ea2);
		\draw[interaction] (xa) -- node[label=right:$\varepsilon_{|\bvec{k}'-\bvec{k}|}v_{|\bvec{k}'-\bvec{k}|}$] {}(xb);
		\draw[electron1] (eb1) -- node[label=above left:$\left|\mathbf{k}_1\mathrm{,}\sigma_1\right\rangle$, xshift=10pt, yshift=5pt] {}(xb);
		\draw[electron2] (xb) -- node[label=above right:$\left|\mathbf{k}_1+\mathbf{k}-\mathbf{k}'\mathrm{,}\sigma_1'\right\rangle$, xshift=0, yshift = 5pt] {}(eb2);
	\end{tikzpicture}	

 \caption{Coulomb scattering vertex connecting electronic initial states  $|\bvec{k},\sigma\rangle$, $|\bvec{k}_1,\sigma_1\rangle$ to final states $|\bvec{k}',\sigma'\rangle$,  $|\bvec{k}_1+\bvec{k}-\bvec{k}',\sigma_1'\rangle$. The spin dependence is due to the influence of spin-orbit coupling in the initial and final states, cf.\ Eq.~\eqref{eq:W-sigma}. The interaction line is the statically screened Coulomb potential.}   
  \label{fig:electron-electron_scattering_image}
\end{figure}

\subsection{Electron-Phonon and Magnon-Phonon Interactions}

We describe electronic energy relaxation processes that occur due to the electron-phonon interaction using a relaxation-time approximation
\begin{equation}
    \frac{\partial }{\partial t } n_{\textbf{k},\sigma}\Big|_{\text{e-p}}=- \frac{ n_{\textbf{k},\sigma}(t) - n^{(\text{eq})}_{\textbf{k},\sigma}}{\tau_{\text{e-p}}},
    \label{df_ep}
\end{equation}
where
\begin{equation}
    n_{\bvec{k},\sigma}^{(\text{eq})}=f_{\text{FD}}\big(\epsilon_{\bvec{k},\sigma}-\mu(n_{\sigma},T_{\text{L}})\big)
\label{eq:n-eq}
\end{equation} 
are Fermi-Dirac distributions that are computed from the dynamical electron distributions $n_{\textbf{k},\sigma}(t)$. The Fermi-Dirac distributions replicate the instantaneous electron density~$n_{\sigma}$ in each band $\sigma$, but are determined with the equilibrium temperature of the lattice, which is kept fixed throughout the dynamics at the lattice temperature $T_{\text{L}} = 300$\;K. This approximation, which introduces the relaxation time $\tau_{\text{e-p}}$ as its characteristic time scale, is made here because electron-phonon scattering occurs on a longer timescale than both electron-electron and electron-magnon scattering so that a microscopic treatment is less important. Note that the contribution~\eqref{df_ep} by itself does not change the total spin of the electronic system, i.e., it does not include spin-flips due to electron-phonon scattering in connection with spin-orbit coupling.

In the numerical results presented below it turns out that the coupling of the magnons to phonons also plays a significant role. In order to describe the effect of such a coupling we also use a relaxation-time approximation, which here takes the form
\begin{equation}
    \frac{\partial }{\partial t } N_{\textbf{q}}\Big|_{\text{m-p}} =- \frac{N_\textbf{q}(t) - N^{(\text{eq})}_\textbf{q}}{\tau_{\text{m-p}}}\ .
    \label{df_mp}
\end{equation}
We again make a bath assumption for the phonons, i.e., we use an equilibrium Bose distribution $N_\textbf{q}^{(\text{eq})}$ at the lattice temperature. The characteristic time scale is set by the magnon-phonon relaxation time~$\tau_{\text{m-p}}$. Equation~\eqref{df_mp} describes the thermalization of the magnon system via processes that change the total number of magnons, i.e., both total magnon energy and total magnon angular momentum are changed. 

The equations for the dynamics of electrons and magnons, which form the basis of the results presented in Sec.~\ref{Results_and_discussion}, are Eqs.~\eqref{e-m_n_k_up}-\eqref{df_em},~\eqref{df_ee},~\eqref{df_ep} and~\eqref{df_mp}. As the scattering processes described by Eqs.~\eqref{e-m_n_k_up}--\eqref{df_em} and~\eqref{df_ee} conserve the total number of electrons, it is important to also ensure this conservation for the numerical solution because differences in the distributions, such as the polarization $P$ defined in~\eqref{def_spin_polarization} play an important role. In these quantities, small deviations from a complete number-density conservation may potentially yield large numerical errors.

\subsection{Electronic Band Structure \label{subsec:energy-magnetization}}

\begin{figure}[h!]
   \centering
    \includegraphics[width=0.45\textwidth]{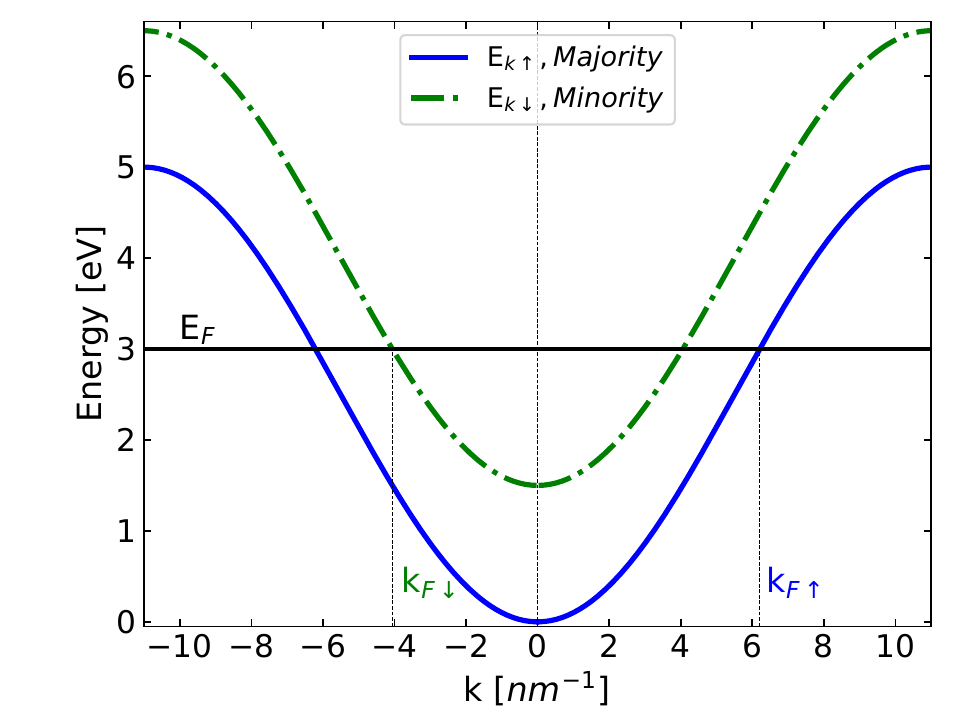}
    \caption{Band dispersion for spin-up (majority) and spin-down (minority) electrons, with a lattice constant $a_0$ of 2.86\,\AA\;and a band width $4\gamma$ of 5\,eV~\cite{edwards_paramagnetic_1982}. Using an effective Hubbard repulsion energy $U_{\text{eff}}=2.7$\,eV, the electron density is chosen such that the self-consistently calculated ground state results in a spin splitting~$\Delta$ of 1.5 eV~\cite{sakoh1975magnetic}.}   
  \label{band_dispersion_energy}
\end{figure}

The description of scattering processes has been fairly general so far and needs to be augmented by specifying the electronic Bloch states connected in these transitions. As a simple model of electronic states in a ferromagnet we consider a band structure with two bands that are separated by a Stoner splitting. For definiteness, we refer to the majority electrons as spin-up electrons and the minority carriers as spin-down, as shown in Fig.~\ref{band_dispersion_energy}.
We choose a modified tight binding (TB) band structure of the form 
\begin{equation}
    \begin{split}
        \epsilon_{\bvec{k},\uparrow} &= \epsilon_0 - 2\gamma \cos(k a_0) \\
        \epsilon_{\bvec{k},\downarrow} &= \epsilon_{\bvec{k},\uparrow} + \Delta 
    \end{split}
    \label{eq:spherical-band}
\end{equation}
where $\gamma$ is the hopping energy that controls the band width, which is in 1D equal to $4\gamma$, and $\Delta$ is the Stoner spin splitting energy. While this result applies to a one-dimensional lattice, we use it here with the modulus $k=|\bvec{k}|$ of a three-dimensional vector. 

There have been very few attempts at the numerically extremely challenging computation of the electron-electron scattering dynamics for multiple bands in the whole Brillouin zone~\cite{mueller_ab_2016}. The results of the momentum and band-resolved calculations are in qualitative agreement with those obtained by using an effective density of states~\cite{mueller_feedback_2013}. Together with the results of Ref.~\cite{stiehl_role_2022} discussed in the Introduction, this suggests that it is not necessary to include the full anistotropic multi-band structure to model the demagnetization dynamics, and we will use the simple $|\bvec{k}|$ band structure~\ref{eq:spherical-band} in the following.

The Stoner splitting is given by
\begin{equation}
    \Delta = U_{\text{eff}}(n_\uparrow - n_\downarrow)
    \label{eq:Stoner-Delta}
\end{equation}
where $n_\sigma$ is the number of electrons per unit cell in the majority/minority channel, respectively, 
\begin{equation}
    n_\sigma := \frac{1}{\mathcal{N}}\sum_{\bvec{k}} n_{\bvec{k}, \sigma},
    \label{def_band_density}
\end{equation}
and $U_{\text{eff}}$ an effective Hubbard electrostatic repulsion energy~\cite{white1968magnon}. 
The Stoner splitting $\Delta$ is self-consistently calculated for the thermal-equilibrium case. We will also discuss the influence of a time-dependent splitting~$\Delta(t)$ which is calculated from the instantaneous values of $n_\sigma = n_\sigma (t)$~\cite{mueller_feedback_2013}. For the electronic states and equilibrium distributions shown in Fig.~\ref{band_dispersion_energy}, we obtain a screening parameter of~$\kappa = 5.1\,\text{nm}^{-1}$.

% ==============================
% ========== RESULTS ===========
% ==============================
\section{Results  and discussion}
\label{Results_and_discussion}

\subsection{Spin Polarization, Magnetization and Their Connection}

Electrons and magnons can, in principle, contribute to the total magnetization of the system, as they both carry angular momentum. The question how to correctly account for this influence on the different effects that are used to experimentally characterize the out-of-equilibrium magnetization is complicated. We follow here Refs.~\cite{beens_s-d_2020,beens_modeling_2022} and combine the spin polarization with the magnon angular momentum to obtain a measure of the total transient magnetization of the system. Switching to dimensionless quantities, we will call the magnetization contribution of the electronic system \emph{spin polarization}~$P$, which is defined as
\begin{equation}
    P = \frac{ n_\uparrow (t) - n_\downarrow (t)}{2}
\label{def_spin_polarization}
\end{equation}
where $n_{\uparrow/\downarrow}$ are the number of majority and minority electrons per unit cell defined in Eq.~\eqref{def_band_density}. 
The contribution of the magnons is the \emph{magnon magnetization} 
\begin{equation}
    M = -\delta N(t)\equiv -[N(t) - N^{(\text{eq})}],
\label{def_magnon_magnetization}
\end{equation}
which is given by the change in the number of magnons 
\begin{equation}
     N(t) = \sum_\bvec{q} N_{\bvec{q}}(t)
     \label{def_n}
\end{equation}
with respect to that in equilibrium $N^{(\text{eq})}$.  The total magnetization $m$, is then determined as 
\begin{equation}
    m = M + P.
    \label{def_magnetization}
\end{equation}

\subsection{Electronic Distribution Functions}

In this and the following subsections we present numerical results for the coupled electron and magnon dynamics due to the contribution of spin-flip electron-electron and electron-magnon (e-e-m) scattering processes combined with electron-phonon and magnon-phonon relaxation processes discussed in Sec.~\ref{methodology}. We first focus on the energy resolved distributions of electrons and magnons computed from Eqs.~\eqref{e-m_n_k_up}--\eqref{df_em},~\eqref{df_ee},~\eqref{df_ep} and~\eqref{df_mp}. From these, ensemble averaged quantities are calculated, in particular the magnetization, for which the physics is discussed starting in Sec.~\ref{subsec:spin-polarization}.

In an experiment, demagnetization is triggered by excitation with an ultrashort pulse that deposits energy in the electronic system. In order to capture this essential characteristic of an optical-pulse excitation, we will consider here an instantaneous excitation in the electronic system, which replaces the equilibrium distribution functions in each spin channel with hot Fermi-Dirac distributions at an electronic temperature of 2000\,K of the same density. This process instantaneously creates ``hot'' electrons without changing the equilibrium spin polarization, i.e., the band structure, specifically the splitting~$\Delta$, is not changed by the excitation. The resulting ``hot'' distribution functions are characterized by a change of the chemical potential of roughly 27\;meV and a spin accumulation of $\zeta =\mu_\uparrow - \mu_\downarrow = 0.6$\,meV, where the chemical potentials refer to the ``hot'' Fermi-Dirac distributions with the same spin polarization as the equilibrium system. Below, we will also use fits by Fermi-Dirac and Bose distributions, respectively, in order to describe the energy and spin polarization by time dependent intensive quantities $T(t)$ and $\mu_s(t)$. However, the underlying distributions are computed dynamically and never assumed to be exact Fermi-Dirac or Bose distributions.

We employ relaxation-time approximations~\eqref{df_ep}, \eqref{df_mp} for the interaction of electrons and magnons  with phonons and choose values of $\tau_{\text{e-p}}=1$\,ps~\cite{groeneveld1992effect, mongin2019ultrafast, voisin2001ultrafast} and $\tau_{\text{m-p}}=10$\,ps~\cite{liu2017magnon}, respectively. The magnon-phonon process is included here to ensure that the system eventually reaches a complete equilibrium. The numerical value of the magnon-phonon relaxation time is such that the demagnetization process at early times remains essentially unaffected. The importance of the relaxation contributions will be discussed in more detail below.

Since we aim at describing a metallic system at high densities, only electrons around and above the Fermi energy play an important role in the dynamics. Their contribution to the $k$-resolved carrier distribution is 
\begin{equation}
     \delta n_{\bvec{k},\sigma} \equiv n_{\bvec{k},\sigma}-f_{\text{FD}}(\epsilon_{\bvec{k},\sigma}- \mu)
    \label{delta_n}
\end{equation}
where we subtract the Fermi-Dirac distribution function for the electrons at equilibrium before excitation, including the equilibrium chemical potential~$\mu$. 

\begin{figure}[t!]
   \centering
    \begin{tikzpicture}
	     \subfigure[]{
	       \node [inner sep=0pt,above right] 
	       {\includegraphics[width=0.49\textwidth, valign=t]{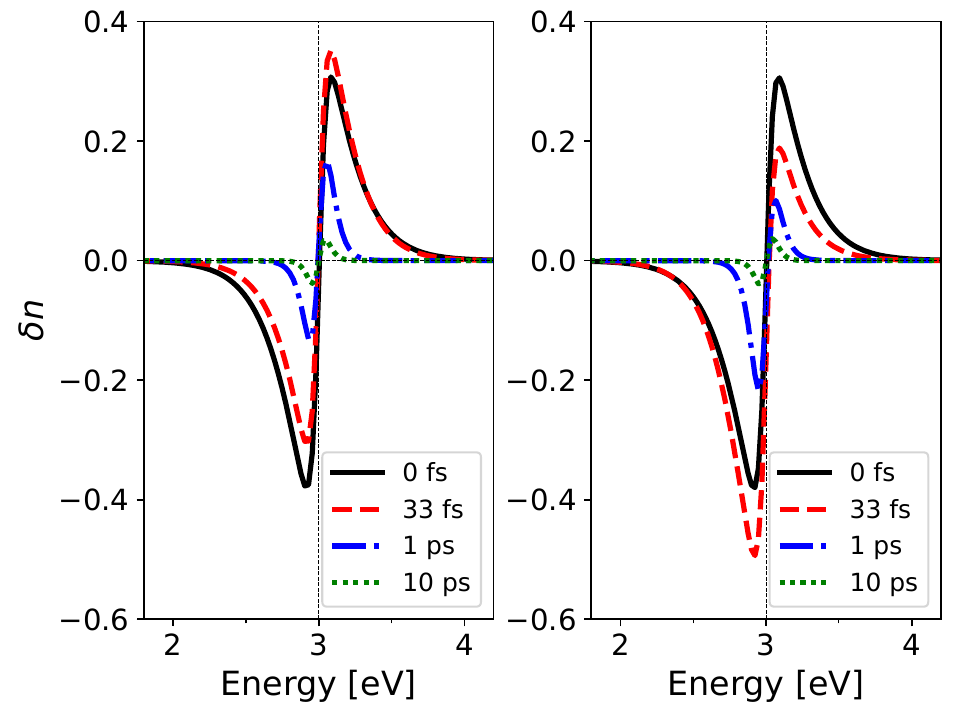}}; 
 	    }
	\node[anchor=west] (a) at (1.5, 6.1) {(a)} ;
	\node[anchor=west] (a) at (5.7, 6.1) {(b)};
      \end{tikzpicture}
    \caption{Energy-resolved change $\delta n_{\bvec{k},\sigma}$ of majority (a) and minority (b) electron distribution functions for different times due to electron-electron and electron-magnon scattering. An instantaneous excitation at $t=0$ is assumed and the change in the distributions is computed with respect to their equilibrium values. Electron-phonon and magnon-phonon interacions are included as relaxation times. The red curve shows that minority electrons are initially scattered into empty majority states.}
  \label{electron_distribution_f}
\end{figure}

Figure~\ref{electron_distribution_f} shows the changes in majority $\delta n_{\bvec{k}\uparrow}$ (a) and minority  carriers  $\delta n_{\bvec{k}\downarrow}$ (b) at different times. The instantaneous excitation at $t=0$ leads to a dipolar shape (black lines), which indicates a redistribution of electrons from below to above the Fermi edge.
The dashed red curve in Fig.~\ref{electron_distribution_f} shows that minority electrons, i.e., carriers in states with energies above $E_{\text{F}}$ are dominantly scattered into majority states ($e_{\downarrow} \rightarrow e_{\uparrow}$). If we refer to carriers below the Fermi level, $E<E_{\text{F}}$, as holes $(h)$, then majority holes are scattered to minority states ($h_{\uparrow} \rightarrow h_{\downarrow}$) at very early times during the dynamics. The snapshot at 33\,fs has been selected as it shows the maximum of this effect, i.e., when the electronic spin polarization reaches a \emph{maximum}, highlighting the deviation from equilibrium in the electronic dynamics. 

The increase in spin polarization is an important characteristic of electron-magnon scattering, as discussed in Sec.~\ref{subsec:spin-polarization} in more detail. At later times, cf.~the blue dash dotted and the green dotted lines in Fig.~\ref{electron_distribution_f}, the $\delta n$s decay as the carrier distributions approach their equilibrium values on the time scale set by the electron-phonon and electron-magnon relaxation times. 

\subsection{Coupled Dynamics of Electrons and Magnons}

\begin{figure}[t!]
   \centering
   \begin{tikzpicture}
	 \subfigure[]{
	   \node [inner sep=0pt,above right] 
	   {\includegraphics[width=0.5\textwidth, valign=t]{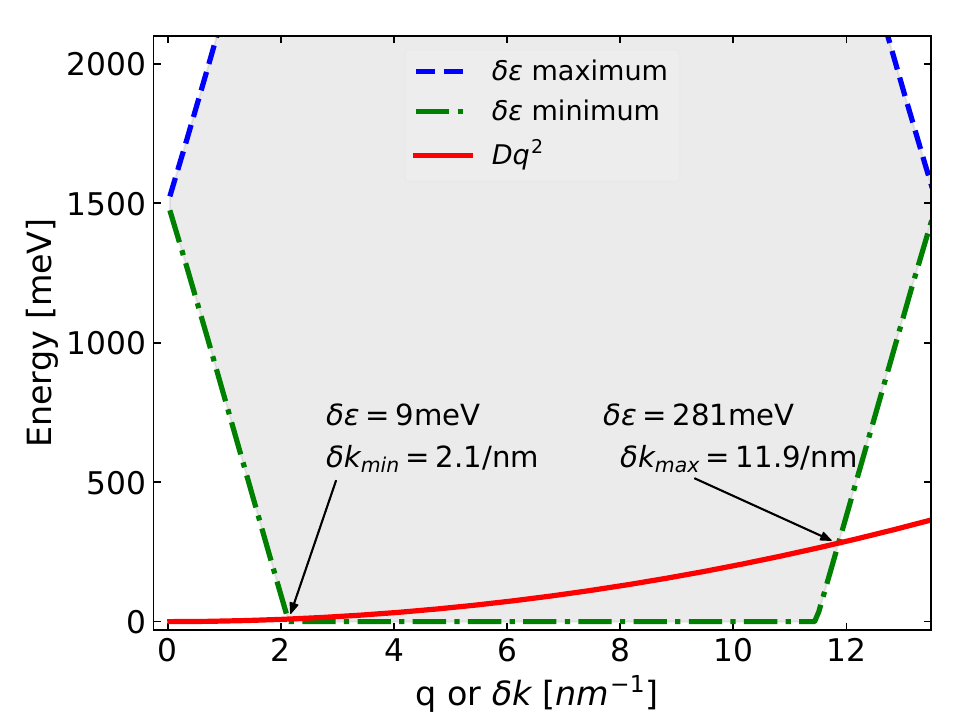}};
 	   \label{stoner_continuum_sfig_a} }
      \subfigure[]{
          \node [inner sep=0pt,below right] 
          {\includegraphics[width=0.5\textwidth, valign=t]{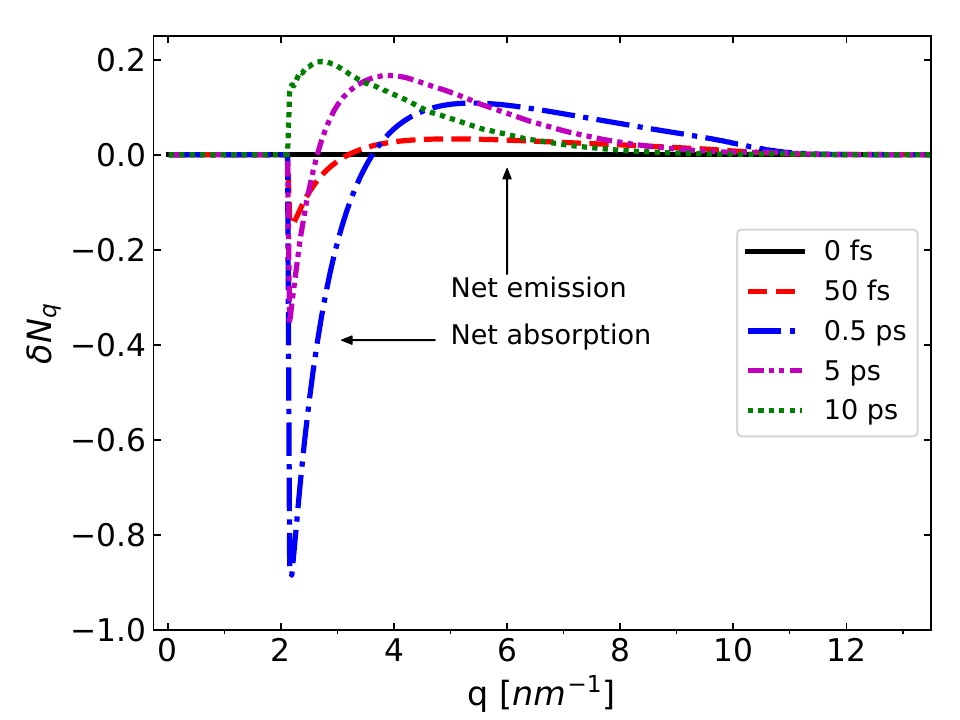}};
          \label{stoner_continuum_sfig_b} }
      \node[anchor=west] (a) at (-0.1, 6.2) {(a)} ;
      \node[anchor=west] (b) at (-0.1, -0.5) {(b)};
   \end{tikzpicture}
   \caption{(a) Dispersion of electronic spin-flip transitions~$\delta\epsilon(\delta k)$ and magnons~$\omega(q)$, cf. Eq.~\eqref{magnon_energy} with $D=2\;$meV\;nm$^2$. The curves overlap between $q_{\text{min}}= 2.1/\text{nm}$ and $q_{\text{max}}= 11.9/\text{nm}$. (b) Change in magnon distribution functions with respect to equilibrium at different times due to scattering with electrons in a band structure with constant exchange splitting. The magnon distributions are changed only in the $q$ region that overlaps with the Stoner continuum.}   
   \label{stoner_continuum}
\end{figure}

This subsection focuses on  the non-equilibrium magnon distributions as obtained from evaluating Eq.~\eqref{df_em}, which sensitively depend on the available scattering phase space for electronic transitions between minority and majority states. Such transitions $|\bvec{k},\uparrow\rangle\to|\bvec{k}+\delta\bvec{k},\downarrow\rangle$ due to electron-magnon scattering with momentum transfer $\delta\bvec{k}$ lead to an electronic energy change 
\begin{equation}
    \delta \epsilon  = \epsilon_{\bvec{k} + \delta\bvec{k}} - \epsilon_{\bvec{k}} + \Delta .
    \label{stoner_excitation_equation}
\end{equation}
in the band structure of Fig.~\ref{band_dispersion_energy}. The relation between energy transfer and momentum change~\eqref{stoner_excitation_equation} is illustrated in Fig.~\ref{stoner_continuum_sfig_a}. The shaded area denotes those $\delta\epsilon (\delta\bvec k)$ values achievable for spin-flip transitions connecting occupied states $|\bvec{k}|\leq k_{F,\uparrow}$ with unoccupied minority states $|\bvec{k}+\delta\bvec{k}|\leq k_{F,\downarrow}$. 

Figure~\ref{stoner_continuum_sfig_a} also shows the magnon dispersion~\eqref{magnon_energy} and its crossing points with allowed spin flip transitions. In the dynamical calculations the non-equilibrium distributions will lead to slightly different picture, as the sharp edges of the shaded area in Figure~\ref{stoner_continuum_sfig_a} arise from the $T=0$\,K Fermi functions, but the plot  qualitatively shows that electronic spin-flip transitions can only couple to magnons with wave vectors $q$ in a limited range of k vectors. In the figure, these points are indicated by $\delta k_{\text{min}}$ and $\delta k_{\text{max}}$.

Figure~\ref{stoner_continuum_sfig_b} plots the computed change in the magnon distribution $\delta N_{q}= N_q(t)-N_q^{(\text{eq})}$. We assume that magnons are not directly excited, so that the black line in Fig.~\ref{stoner_continuum_sfig_b} at $t=0$ is flat, $\delta N_q(t=0)=0$. After the instantaneous excitation process, only magnons in the wave-vector range between $\delta k_{\text{min}}$ and $\delta k_{\text{max}}$ are absorbed and emitted. This is the range where the magnon dispersion overlaps with the Stoner continuum. In more detail, during the early stages of the coupled dynamics, shown here for 50\,fs and 0.5\,ps, magnons at long wavelengths, i.e., $q$ near $q_{\text{min}}$ are predominantly absorbed, whereas those at shorter wavelengths, i.e., $q$ between $q\simeq 3$\,nm$^{-1}$ and $q_{\text{max}}$, are predominantly emitted. The absorption for small $q$ is much weaker than the emission for large $q$, because the total change in magnon density is calculated according to Eq.~\eqref{def_n} as 
\begin{equation}
     N(t) = \frac{\Omega}{2\pi^2}\int q^2 N_{q} dq
\end{equation}
including the weight factor of $q^2$. At longer times, shown here for  5\,ps and 10\,ps, the absorption range  of $\delta N_q$ slowly shrinks until there is no trace left while the peak resulting from effective magnon emission shifts towards $q_{\text{min}}$. Even for 10\,ps the magnon distributions still have not returned to equilibrium, and the same is true for the electronic distributions in Fig.~\ref{electron_distribution_f}.

We stress that magnon decay processes beyond the electron-magnon coupling are taken into account in the dynamics of the magnon distribution functions via a phenomenological relaxation term that is intended to mimic magnon-phonon interactions. This is not the same as the Landau damping of magnons, for which one contribution is the decay into single-particle Stoner excitations, which is covered by the electron-magnon coupling. The latter processes occur in the overlap region of the magnon dispersion with the shaded area in Figure~\ref{stoner_continuum_sfig_a}. This and other contributions to magnon damping are studied in an ab-initio fashion, for instance, in Ref.~\cite{paischer2024correlations}.

\begin{figure}[t!]
   \centering
    \includegraphics[width=0.48\textwidth]{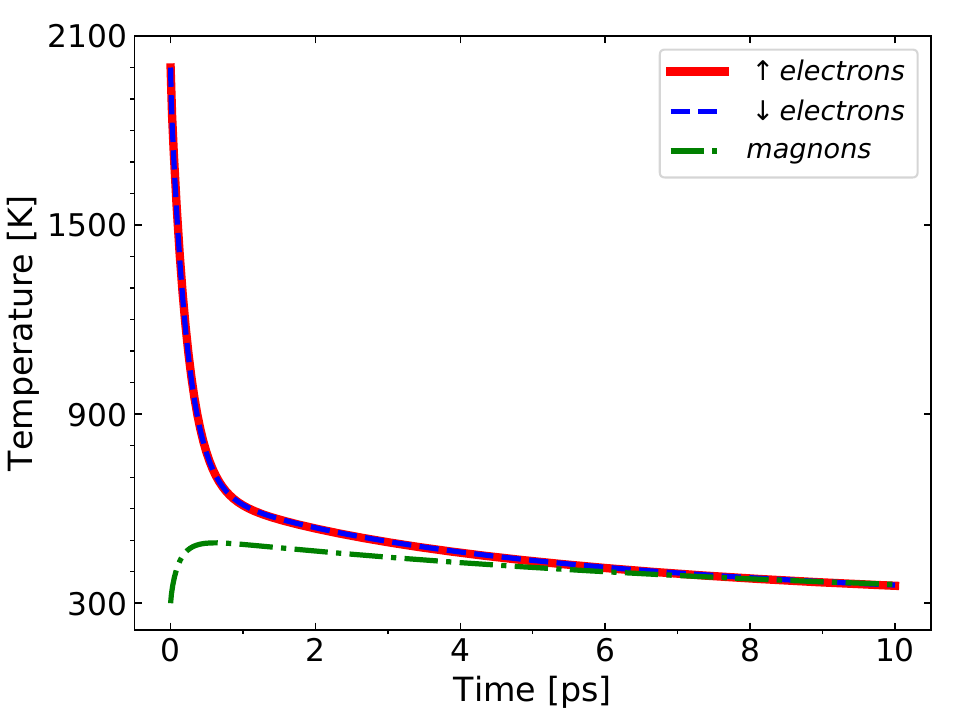} 
    \caption{Temperature of electrons and magnons during scattering dynamics. Only electrons are excited at 2000\,K while the magnons are kept at the lattice temperature 300\,K. }   
  \label{temperature}
\end{figure}

Figure~\ref{temperature} plots the effective temperatures of electrons and magnons after the instantaneous excitation. We stress that these effective temperatures are those of equilibrium distributions with the same particle number and energy densities as those of the dynamical distributions, and show the energy exchange between electrons, magnons, and the lattice. They are determined by fits to the calculated dynamical distributions. At all times, the effective temperatures for spin-up and spin-down electrons are almost equal, due to the excitation condition and the fast electron-electron scattering. At times shorter than a picosecond, most of the energy is exchanged between electrons and magnons due to their interaction as the instantaneous excitation opens up scattering phase space. The magnons take about 0.5\,ps until their maximum effective temperature of roughly 500\,K is reached. At this point, the blue dash dotted line in Fig.~\ref{stoner_continuum_sfig_b} shows that the deviation of the magnon distribution function from its equilibrium is most pronounced. After approximately 7\,ps, the energy transfer from electrons to magnons has slowed down and a quasi-equilibrium situation is reached, in which the common electron and magnon temperature still differs from that of the lattice. Complete thermal equilibrium, at which all temperatures are equal to 300\,K, is only approached on the time scale of the magnon-phonon relaxation. 

\subsection{Spin Polarization and Magnetization Dynamics\label{subsec:spin-polarization}}

\begin{figure}[t!]
   \centering
   \begin{tikzpicture}
	 \subfigure[]{
	   \node [inner sep=0pt,above right] 
	   {\includegraphics[width=0.48\textwidth, valign=t]{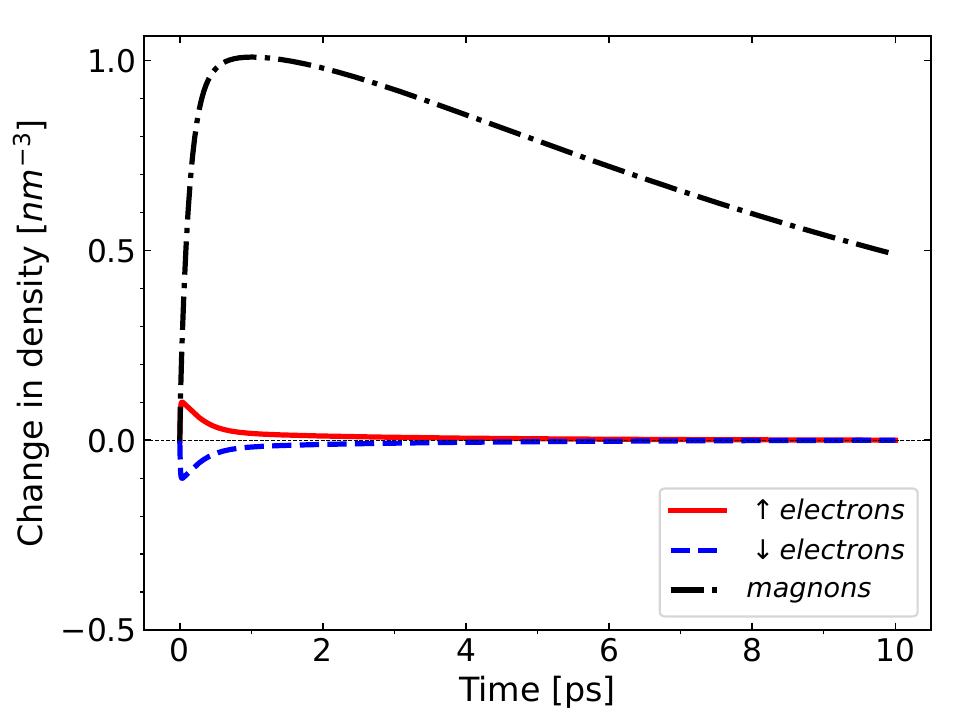}};
 	   \label{figure_7_sfig_a} }
      \subfigure[]{
          \node [inner sep=0pt,below right] 
          {\includegraphics[width=0.48\textwidth, valign=t]{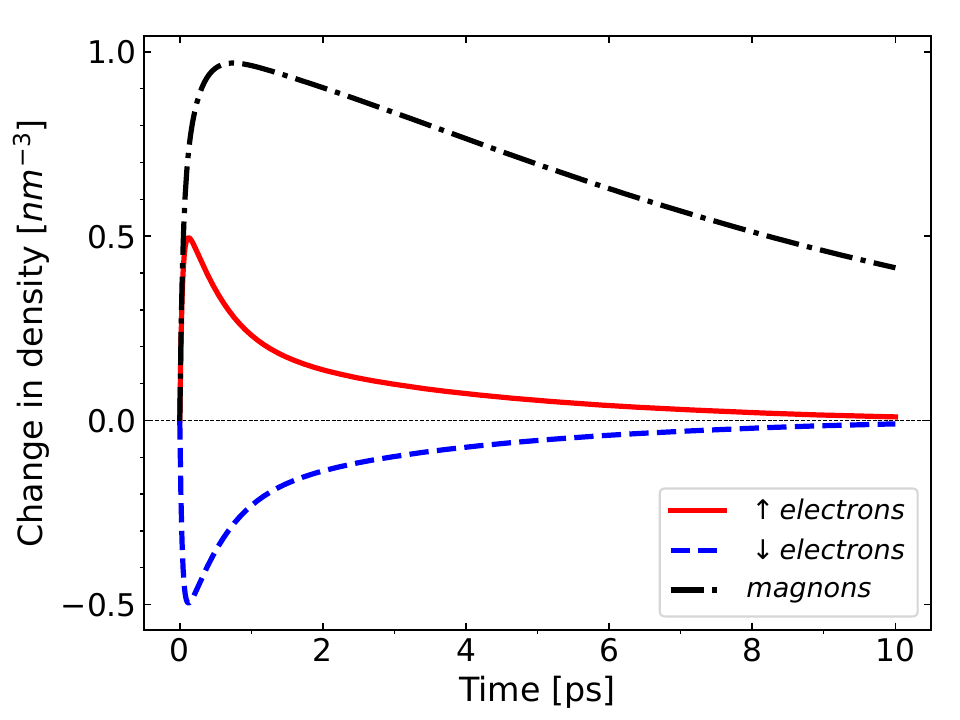}};
          \label{figure_7_sfig_b} }
      \node[anchor=west] (a) at (-0.1, 6.3) {(a)} ;
      \node[anchor=west] (b) at (-0.1, -0.28) {(b)};
   \end{tikzpicture}
   \caption{Density change for spin up (solid line), spin down (dashed blue line) electrons and magnons due to electron-electron and electron-magnon scattering with electron-phonon and magnon-phonon coupling treated by relaxation rates; (a) uses a constant spin splitting~$\Delta$ and (b) a dynamical~$\Delta(t)$.}   
   \label{fig-7}
\end{figure}

After the characterization of the electron and magnon dynamics at the level of microscopic energy resolved distributions, we now turn to the macroscopic quantities, in particular the magnetization dynamics.
In Fig.~\ref{fig-7} we investigate the different contributions to the magnetization as defined in Eqs.~\eqref{def_spin_polarization}--\eqref{def_magnetization}. We consider in this section the cases of a fixed splitting $\Delta$ and a dynamically calculated~$\Delta(t)$~\cite{mueller_feedback_2013}, which changes the electronic quasi-particle states. At the end, we compare both scenarios, but the main conclusion can already be drawn by considering the case with constant $\Delta$, which we discuss in detail first.

Figure~\ref{fig-7} plots the change in the electronic $\delta n_{\uparrow/\downarrow}$ and magnon densities $\delta N$. The changes of the electronic densities have the same magnitude but opposite sign, so that the electronic spin-polarization is just $P = \delta n_{\uparrow}$, cf.~Eq.~\eqref{def_spin_polarization}, whereas the magnon magnetization $M$ is the change in magnon density plotted in Fig.~\ref{fig-7} with opposite sign. At very early times, the electronic spin polarization \emph{increases}  as magnons are created, which is the net effect when integrating the change in the magnon distribution $\delta N_q$ as shown in Fig.~\ref{stoner_continuum_sfig_b} over all wave vectors $q$, cf. Eqs.~\eqref{def_magnon_magnetization}--\eqref{def_n}. In the case of a constant gap as shown in Fig.~\ref{figure_7_sfig_a} the change in magnon density is almost 10 times larger than the density of the electronic spins flipped, or, equivalently, the decrease in $M$ is 10 times larger than the increase in $P$. The effect of the scattering processes is, therefore, to produce a change in the magnon magnetization that is an order of magnitude larger than the change in the spin polarization. 

This imbalance is an important result, because electron-magnon scattering alone leads to changes in magnon density that \emph{exactly equal} the electronic spin-polarization change, because the angular-momentum conservation is implicitly included in Eqs.~\eqref{eq:V-e-m}--\eqref{df_em}. The large discrepancy between $M$ and $P$ in Fig.~\ref{figure_7_sfig_a} can be explained by the \emph{interplay} of electron-magnon with EY-type electron-electron scattering as follows. On the timescale on which most of the energy is transferred from the electronic to the magnon system, as shown in Fig.~\ref{temperature}, electron-magnon scattering creates magnons in a process that transfers angular momentum into the electronic system, as sketched in Fig.~\ref{fig:electron-magnon_scattering_image}. The electron-magnon scattering processes thereby increase the spin polarization or, equivalently, lead to a non-equilibrium spin accumulation $\zeta =\mu_\uparrow(t)-\mu_{\downarrow}(t)$. We reiterate that the chemical potentials $\mu_\sigma(t)=\mu\big(n_\sigma(t),T_\sigma(t)\big)$ and effective temperatures $T_\sigma(t)$ refer to Fermi-Dirac distributions in each band that have the same density and energy density as the non-equilibrium distributions $n_{\bvec{k},\sigma}(t)$. The non-equilibrium spin accumulation in the electronic systems is the driving force for EY-type electron-electron scattering~\cite{mueller_driving_2011,mueller_feedback_2013}, which effectively relaxes the spin polarization and, in turn, frees up scattering phase space in the electronic system for spin-flip transitions that create more magnons. The two processes can support each other until the maximum magnon density is reached. At that time, the electronic system has cooled down so much that the electron-magnon scattering has only little scattering phase-space left. 

It is interesting to compare the combination of scattering processes with the electron-electron scattering of EY-type acting alone, i.e., without magnons~\cite{krauss2009ultrafast,mueller_feedback_2013}. In this case we would find a \emph{decrease} of the spin-polarization compared to its equilibrium value, because EY-like scattering relaxes the spin accumulation that is created by the instantaneous heating process with a sign opposite to that driven by electron-magnon scattering. The absolute value of the change in spin polarization would be considerably smaller (not shown) than what is shown in Fig.~\ref{figure_7_sfig_a} for the band structure and excitation conditions of the present paper.

\begin{figure}[tb!]
	\centering
	\begin{tikzpicture}
		\subfigure[]{
			\node [inner sep=0pt,above right] 
			{\includegraphics[width=0.5\textwidth, valign=t]{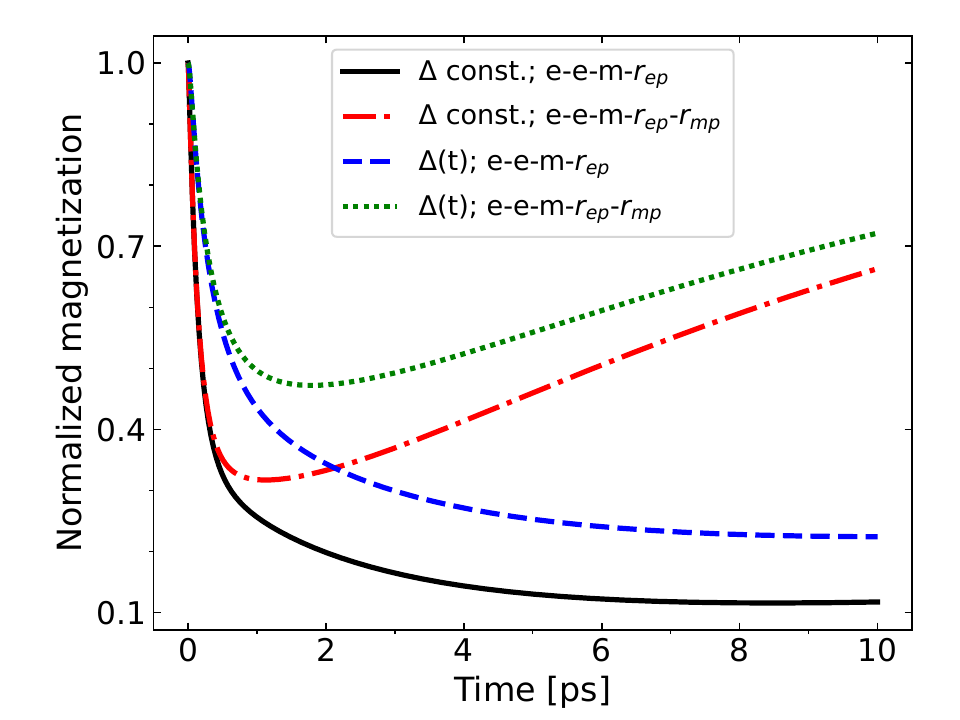}};
			\label{change_magnetization_sfig_a} }
		\subfigure[]{
			\node [inner sep=0pt,below right] 
			{\includegraphics[width=0.5\textwidth, valign=t]{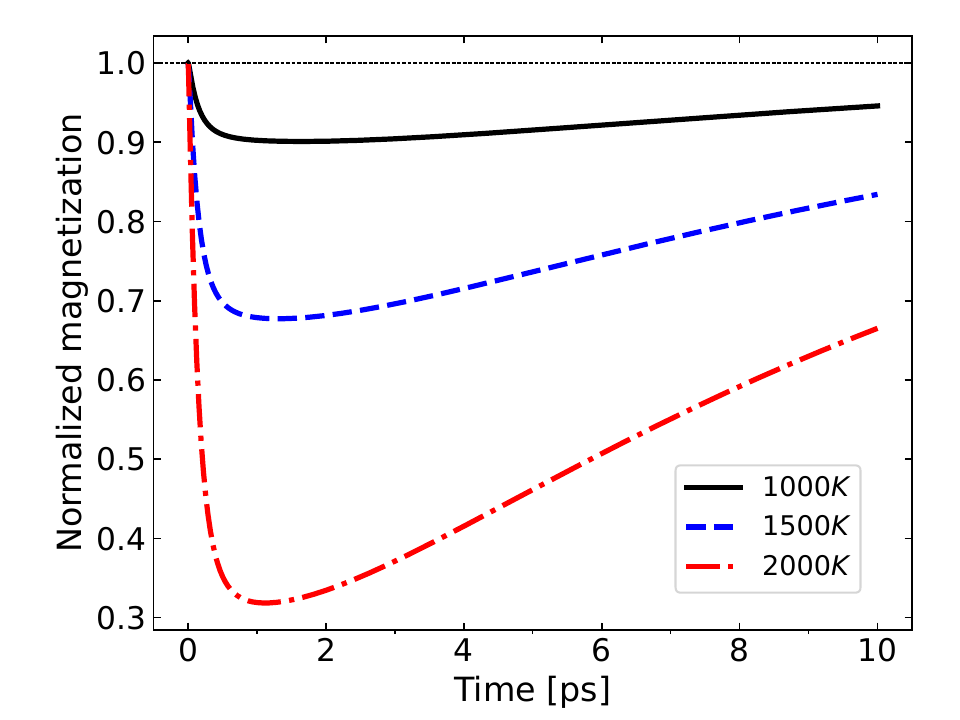}};
			\label{change_magnetization_sfig_b} }
		\node[anchor=west] (a) at (2.3, 6.0) {(a)} ;
		\node[anchor=west] (b) at (2.3, -0.8) {(b)};
	\end{tikzpicture}
	\caption{Magnetization dynamics $m(t)$ after instantaneous heating including electron-electron and electron-magnon scattering contributions together with electron-phonon relaxation ($r_{\text{ep}}$). (a) Results computed for constant and dynamic splitting $\Delta$ with and without magnon-phonon relaxation ($r_{\text{mp}}$). (b) Magnetization dynamics including all scattering and relaxation contributions for a constant splitting  $\Delta$ and different excitation temperatures.}   
	\label{change_magnetization}
\end{figure}

We now turn to the results of the calculation including the dynamical splitting in Fig.~\ref{figure_7_sfig_b}. Qualitatively the same observations hold true here, but the maximum transient electronic spin polarization is about 5 times larger than that occurring for the constant splitting in Fig.~\ref{figure_7_sfig_a} and appears later at approximately 130\,fs. The dynamical gap, i.e., an instantaneous change of the mean-field~\eqref{eq:Stoner-Delta}, amplifies the change in electronic spin polarization compared with the case of a static gap, as the change in splitting increases the available phase space for whatever process drives the spin-changing transitions in the electronic system. In Ref.~\cite{mueller_feedback_2013}, where the case of a pure EY-like scattering processes with electrons and phonons was considered, this interplay of the scattering processes with the dynamical gap leads to a stronger decrease of the spin polarization after an instantaneous heating.

Figure~\ref{change_magnetization} shows the dynamics of the magnetization $m(t)$ normalized to the electronic polarization  in equilibrium. As before, we include electron-phonon and magnon-phonon interactions as relaxation processes, but ignore the magnon-phonon relaxation in some of the results to highlight its impact. While Fig.~\ref{change_magnetization_sfig_a} focuses on the impact of magnon-phonon relaxation and on the difference between a constant and dynamic Stoner splitting, Fig.~\ref{change_magnetization_sfig_b} illustrates the influence of the excitation strength by varying the temperature of the initial hot Fermi-Dirac distribution.

We first consider those results in Fig.~\ref{change_magnetization_sfig_a} that include only electron-phonon relaxation processes. In this case, the magnetization steadily decreases, with a stronger demagnetization in the constant-gap case. The total magnetization $m=M+P$ is quenched less for the case of the dynamical gap. The difference is almost exclusively due to the contributions of the electronic spin polarization~$P$, cf.~Fig.~\ref{fig-7}.
For longer times, the magnetization approaches a constant value, which is a result of ``frozen out'' magnons. These are non-equilibrium magnons created by the electron-magnon scattering at earlier times at $q$ values that are not accessible any more due to a lack of electronic scattering partners in the equilibrated electron distributions. It seems that, in order for the magnons to relax, a direct coupling of phonons to magnons is necessary, which led us to include a magnon-phonon relaxation term at the outset.

The results in Fig.~\ref{change_magnetization_sfig_a}  that include such a magnon-phonon relaxation contribution resemble the typical shape of demagnetization dynamics usually observed experimentally where an ultrafast quenching of the magnetization is followed by a remagnetization on a 10 ps time scale. It is evident that a dynamical gap reduces the minimal magnetization due to a more pronounced increase in spin polarization. 

In Fig.~\ref{change_magnetization_sfig_b} we study the influence of the energy deposited in the electronic system on the demagnetization process.  We find that a smaller quenching of about $30\%$ and $10\%$ results for 1500\;K and 1000\;K, respectively, while maintaining the same overall shape and time scales of the demagnetization and remagnetization dynamics. The deposited energies per unit cell due to the excitation are 8.8\,meV, 20.9\,meV, 38.1\,meV for 1000\,K, 1500\,K, 2000\,K, respectively, which is on the same order of magnitude as in typical experiments~\cite{essert_electron-phonon_2011,stiehl_role_2022} 

These results for the demagnetization dynamics suggest that non-equilibrium magnons play an important role in the demagnetization process and that different magnon-interaction processes are involved in the demagnetization and remagnetization dynamics. The most important simplifying model assumptions, such as the form of the band structure and excitation processes, can be improved in the future using ab-initio techniques such as those employed in Refs.~\cite{muller_electron-magnon_2019,paischer2023nonlocal}. Eventually it should be possible to obtain the achievable demagnetization for a realistic material excitation, i.e., deposited energy per unit cell, and compare these with experiments. This is likely the most important test any candidate demagnetization mechanism mentioned in Sec.~\ref{sec:demag} has to pass.

\section{Conclusion\label{conclusion}}

We investigated  the dynamics of electrons in ferromagnets on ultrashort timescales due to electron-magnon and spin-flip electron-electron scattering in the framework of a microscopic model. In this model the dominant interactions between electrons and magnons that change the electronic spin involve high-$q$ magnons for which the dispersion crosses the $(q,\omega)$ region where electronic spin-flip single-particle transitions are possible. After an instantaneous excitation process, which opens up electronic scattering phase space, non-equilibrium magnon distributions are created. If one considers the idealized case of only electron-magnon scattering, one obtains an increase of the electronic spin polarization that is \emph{equal} in magnitude but opposite to the decrease of angular momentum due to the emission of magnons. 

The balance between spin polarization and magnon magnetization is changed when Elliott-Yafet like spin-flip scattering is included, which is the realistic case for ferromagnetic transition metals. This spin-flip scattering occurs due to the interplay of spin-orbit coupling and electron-electron scattering; it provides a spin relaxation mechanism for the electronic spin polarization that facilitates a larger demagnetization due to efficient magnon emission. This is because the scattering processes creating the magnons also increase the spin polarization, which is then relaxed by the Elliott-Yafet like spin relaxation. 

The results obtained for the demagnetization due the interplay of electron-magnon scattering and Elliott-Yafet-like spin relaxation processes in the framework of our model seem promising. We presented evidence that this candidate mechanism can lead to a realistic magnitude of the demagnetization for a realistic deposited energy per unit cell. This is likely the most important test any candidate demagnetization mechanism mentioned in Sec.~\ref{sec:demag} has to pass. The numerical evaluation of the interplay of scattering processes can be improved from the one presented here to a more material-realistic band structure in the future, which will make it possible to obtain the achievable demagnetization for excitation conditions that can be directly compared with experiments. 

From our model calculations one can draw the conclusion that ultrafast demagnetization involves mainly the emission of magnons, with only a comparatively small contribution of the spin polarization created in the electronic system. We only obtain a remagnetization behavior in accordance with experimental results if we include a phenomenological magnon-phonon coupling. Without this coupling the remagnetization process due to the combination of electron-magnon and electron-phonon scattering results in a steady state with non-equilibrium magnon distributions that cannot change due to scattering with the equilibrated electron system. This result points to the importance of magnon relaxation and decay processes, which we intend to include into the present approach in future papers.

\begin{acknowledgments}
This work was funded by the Deutsche Forschungsgemeinschaft (DFG, German Research Foundation) through grant No.\ TRR 173-268565370 (projects B03, A08) and Deutscher Akademischer Austauschdienst  (DAAD, German Academic Exchange Service). 
\end{acknowledgments}

\section*{Data Availability}
The data and computer codes are not publicly available. They are available from the authors upon reasonable request.

\appendix

\section{Derivation of Electron-Magnon Scattering Contributions}
\label{appendix}

In this Appendix, we derive the scattering contributions to the equations of motion for the distribution function of electrons and magnons due to electron-magnon interaction using a truncation of the hierarchy of the equations of motion~\cite{haug2004quantum,rossi_theory_2002,kira_semiconductor_2012}.
The equation of motion for electron distributions such as $n_{\textbf{k},\downarrow}=\langle c_{\textbf{k},\downarrow}^\dagger c_{\textbf{k},\downarrow} \rangle$ is given by
\begin{equation}
        i\hbar \frac{\partial}{\partial t}n_{\textbf{k},\downarrow} \Big|_{\text{e-m}}= -\frac{M_{\text{e-m}}}{\sqrt{\mathcal{N}}}\sum_\textbf{q}  \big(\big< a_\textbf{q} c_{\textbf{k},\downarrow}^\dagger c_{\textbf{k}-\textbf{q},\uparrow}\big> -\text{c. c.} \big)
\label{eq:n_e-down_1}
\end{equation}
where electron-magnon correlations of the form $\langle a c^\dagger_\downarrow c_\uparrow \rangle $ appear.

Similarly, the distribution function of magnons is given by
\begin{equation}
	i\hbar\frac{\partial }{\partial t}N_{\bvec{q}}\Big|_{\text{e-m}}=\frac{M_{\text{e-m}}}{\sqrt{\mathcal{N}}}\sum_{\bvec k}\big[
	\langle a_{\bvec q}c_{\bvec k,\downarrow}^{\dagger}c_{\bvec k-\bvec q,\uparrow}\rangle
	- \text{c. c.} \big]	
	\label{eq:N-magnon}
\end{equation}

The electron-magnon correlation function obeys
\begin{widetext} 
\begin{equation}
  i\hbar\frac{\partial}{\partial t}\langle a_{\bvec q}c_{\bvec k,\downarrow}^{\dagger}c_{\bvec k-\bvec q,\uparrow}\rangle
  =-\frac{M_{\text{e-m}}}{\sqrt{\mathcal{N}}}\Big\{\sum_{\bvec k'}\langle c_{\bvec k'-\bvec q,\uparrow}^{\dagger}c_{\bvec k',\downarrow}c_{\bvec k,\downarrow}^{\dagger}c_{\bvec k-\bvec q,\uparrow}\rangle
  +\sum_{\bvec q'}\big[\langle a_{\bvec q}a_{\bvec q'}^{\dagger}c_{\bvec k,\downarrow}^{\dagger}c_{\bvec k-\bvec q+\bvec q',\downarrow}\rangle
  -\langle a_{\bvec q}a_{\bvec q'}^{\dagger}c_{\bvec k-\bvec{\bvec q'},\uparrow}^{\dagger}c_{\bvec k-\bvec q,\uparrow}\rangle\big]\Big\}
\end{equation} 

At the scattering level,  the equation of motion for the electron-magnon correlation is closed by replacing higher-order correlations by products of lower-order correlations, such as
\begin{equation}
	\langle c_{\bvec k'-\bvec q,\uparrow}^{\dagger}c_{\bvec k',\downarrow}c_{\bvec k,\downarrow}^{\dagger}c_{\bvec k-\bvec q,\uparrow}\rangle  =\delta_{\bvec k,\bvec k'}\langle c_{\bvec k'-\bvec q,\uparrow}^{\dagger}c_{\bvec k-\bvec q,\uparrow}\rangle-\langle c_{\bvec k'-\bvec q,\uparrow}^{\dagger}c_{\bvec k,\downarrow}^{\dagger}c_{\bvec k',\downarrow}c_{\bvec k-\bvec q,\uparrow}\rangle
	\simeq\delta_{\bvec k,\bvec k'}n_{\bvec k-\bvec q,\uparrow}(1-n_{\bvec k,\downarrow})
\end{equation}
and, similarly,
	$\langle a_{\bvec q}a_{\bvec q'}^{\dagger}c_{\bvec k,\downarrow}^{\dagger}c_{\bvec k-\bvec q+\bvec q',\downarrow}\rangle\simeq\delta_{\bvec q,\bvec q'}(1+N_{\bvec q})n_{\bvec k,\downarrow}$
With these replacements, the electron-magnon correlation dynamics are determined by
\begin{equation}
    \begin{split}
        i\hbar \frac{\partial}{\partial t}\big< a_\textbf{q} c_{\textbf{k},\downarrow}^\dagger c_{\textbf{k}-\textbf{q},\uparrow}\big> = &
        -( \epsilon_\textbf{k}^\downarrow -\epsilon_{\textbf{k}-\textbf{q}}^\uparrow  -\hbar\omega_\textbf{q} + i\hbar\gamma)  \big< a_\textbf{q} c_{\textbf{k},\downarrow}^\dagger c_{\textbf{k}-\textbf{q},\uparrow}\big>   \\
        & +  \frac{1}{\sqrt{\mathcal{N}}} M_{\text{e-m}} \Big[N_\textbf{q}   n_{\textbf{k}-\textbf{q},\uparrow} (1- n_{\textbf{k},\downarrow}) - n_{\textbf{k},\downarrow} (1+N_\textbf{q}) (1 -n_{\textbf{k}-\textbf{q},\uparrow}) \Big] 
    \end{split}
       \label{Eq_correlations_1}
    \end{equation}
\end{widetext}
where we have included a decay constant~$\gamma$ that approximates the influence of higher-order contributions. This equation has the general form
\begin{equation}
     i\hbar \frac{\partial}{\partial t} X =  - (\Delta E+i\hbar\gamma) ~X + i\hbar\Gamma(t),
    \label{eq:dXdt}
\end{equation}  
which is very similar to the corresponding quantity for phonon-assisted electronic distributions. The formal solution of Eq.~\eqref{eq:dXdt} is given by~\cite{haug2004quantum,baral_re-examination_2016}
\begin{equation}
    X (t) =\frac{1}{i\hbar} \int_{-\infty}^t d\tau ~\Gamma (\tau)  e^{i[\Delta E/\hbar + i\gamma](t-\tau)}.
\end{equation}
 Using the Markov approximation, which assumes that $\Gamma(\tau)$ varies slowly compared to the rotation frequency  $\Delta E/\hbar$ and can be approximated as $\Gamma(t)$, this equation can be analytically solved, yielding 
\begin{equation}
    X(t) = \frac{\Gamma (t)}{\Delta E + i \hbar\gamma}.
\end{equation}
If $\gamma \rightarrow 0$, we have  $ X(t)= \Gamma(t)[\mathcal{P} \frac{1}{\Delta E} -i\pi\delta (\Delta E)] $
where $\mathcal{P}$ denotes the principal value, and one obtains
\begin{widetext}
\begin{equation}
    \begin{split}
        \big< a_\textbf{q} c_{\textbf{k},\downarrow}^\dagger c_{\textbf{k}-\textbf{q},\uparrow}\big>  
        &=   \frac{M_{\text{e-m}}}{\sqrt{\mathcal{N}}} \frac{[N_\textbf{q}   n_{\textbf{k}-\textbf{q},\uparrow} (1- n_{\textbf{k},\downarrow}) - n_{\textbf{k},\downarrow} (1+N_\textbf{q}) (1 -n_{\textbf{k}-\textbf{q},\uparrow})  ] }{( \epsilon_\textbf{k}^\downarrow -\epsilon_{\textbf{k}-\textbf{q}}^\uparrow  -\hbar\omega_\textbf{q}) + i \gamma}  \\
        &= -\frac{i \pi}{\sqrt{\mathcal{N}}} \delta ( \epsilon_\textbf{k}^\downarrow -\epsilon_{\textbf{k}-\textbf{q}}^\uparrow  -\hbar\omega_\textbf{q})  M_{\text{e-m}}[N_\textbf{q}   n_{\textbf{k}-\textbf{q},\uparrow} (1- n_{\textbf{k},\downarrow}) - n_{\textbf{k},\downarrow} (1+N_\textbf{q}) (1 -n_{\textbf{k}-\textbf{q},\uparrow})  ]
    \end{split}
\end{equation}
where we have only retained the imaginary part, as the real part drops out or yields an energy renormalization, which we ignore. Then, with Eq.~\eqref{eq:n_e-down_1}, the EOM for the  dynamics of spin-down electron occupations due to electron-magnon scattering becomes
\begin{equation}
     \frac{\partial}{\partial t}n_{\textbf{k},\downarrow} =\frac{2 \pi}{\mathcal{N}\hbar}  \sum_\textbf{q}  M_\textbf{q}^2 \delta ( \epsilon_\textbf{k}^\downarrow -\epsilon_{\textbf{k}-\textbf{q}}^\uparrow  -\hbar\omega_\textbf{q})[N_\textbf{q}   n_{\textbf{k}-\textbf{q},\uparrow} (1- n_{\textbf{k},\downarrow}) - n_{\textbf{k},\downarrow} (1+N_\textbf{q}) (1 -n_{\textbf{k}-\textbf{q},\uparrow})  ],
     \label{spin_full_em_down}
\end{equation}
The equivalent relation for  spin-up electrons is obtained in exactly the same way.
The magnon distribution dynamics is derived starting from Eq.~\ref{eq:N-magnon}. Similar steps as in the above derivation then lead to 
\begin{equation}
    \frac{\partial}{\partial t}N_{\textbf{q}} = \frac{2 \pi}{\mathcal{N}\hbar}  \sum_\textbf{k}  M_\text{e-m}^2 \delta(\epsilon_\textbf{k}^\uparrow -  \epsilon_{\textbf{k}+\textbf{q}}^\downarrow  + \hbar\omega_\textbf{q} ) [  n_{\textbf{k}+\textbf{q},\downarrow} (1- n_{\textbf{k},\uparrow})(1 + N_\textbf{q })  -  N_\textbf{q}  n_{\textbf{k},\uparrow}(1-n_{\textbf{k}+\textbf{q},\downarrow}) ].
    \label{eq_occupations_magnon_em}
\end{equation}
\end{widetext}

%\bibliography{paper_references.bib}
%\input{electron_magnon_resub.bbl}


\end{document}